**Controllable chirality-induced geometrical Hall effect in a frustrated highly-correlated metal**

B. G. Ueland[1*], C. F. Miclea[1,2], Yasuyuki Kato[1], O. Ayala–Valenzuela[1], R. D. McDonald[1], R. Okazaki[3], P. H. Tobash[1], M. A. Torrez[1], F. Ronning[1], R. Movshovich[1], Z. Fisk[1,4], E. D. Bauer[1], Ivar Martin[1], and J. D. Thompson[1]

[1]*Los Alamos National Laboratory, Los Alamos, NM 87544, USA*

[2] *National Institute of Materials Physics 077125 Bucharest-Magurele, Romania*

[3]*Department of Physics, Kyoto University, Kyoto 606-8502, Japan*

[4]*University of California, Irvine, CA 92697, USA*

[*]*bgueland@lanl.gov*

**Abstract:** A current of electrons traversing a landscape of localized spins possessing non-coplanar magnetic order gains a geometrical (Berry) phase which can lead to a Hall voltage independent of the spin-orbit coupling within the material--a geometrical Hall effect. We show that the highly-correlated metal $UCu_5$ possesses an unusually large controllable geometrical Hall effect at $T$<1.2K due to its frustration-induced magnetic order. The magnitude of the Hall response exceeds 20% of the $\nu$=1 quantum Hall effect per atomic layer, which translates into an effective magnetic field of several hundred Tesla acting on the electrons. The existence of such a large geometric Hall response in $UCu_5$ opens a new field of inquiry into the importance of the role of frustration in highly-correlated electron materials.

Frustration is prevalent in nature, appearing in substances as simple as water ice to systems as complex as neural networks[1,2]. In general, frustration is the result of competing interactions of comparable strengths that can prevent a system from entering a single macroscopic lowest energy configuration. The geometry of certain material structures is particularly prone to frustration, with the resulting degenerate ground states or near-ground states being highly susceptible to small perturbations and the emergence of exotic states of matter[1,2]. For example, frustration imposed by the geometrical arrangement of magnetic moments in electrical insulators induces novel states, such as spin-liquid, multiferroicity, and magnetic monopole excitations in spin ice[1-7]. The absence of itinerant charge and spin degrees-of-freedom in magnetically frustrated insulators simplifies theoretical modelling of the microscopic origin and consequences of that frustration and, consequently, most studies have focussed on non-metallic materials. Nevertheless, geometrical frustration of spin configurations also may have significant consequences for metals, especially in highly-correlated materials with itinerant electrons strongly coupled to localized spins[8]. One of those possible consequences is the emergence of a highly tunable and very large response of itinerant charge carriers to an applied magnetic field.

In an ordinary metal, a Hall voltage $V_{Hall}$ arises perpendicular to both the current $I$ and applied magnetic field $H$ and is related to the transverse resistivity $\rho_{xy}=V_{Hall}/I=R_0H$, where $R_0$ is the Hall coefficient[9]. For a magnetic metal, however, scattering of itinerant electrons off of localized spins can produce additional contributions to $\rho_{xy}$, typically proportional to the magnetization $M$ and powers of the longitudinal resistivity $\rho_{xx}=V_{xx}/I$, resulting in so-called anomalous Hall effects described by Karplus-Luttinger[10] and skew scattering theories[9] that are discussed in the Supplementary Methods section in the Supplementary Information. Additionally, if frustration were to induce spin order in a non-coplanar structure, then itinerant electrons traversing the spin texture with local spin chirality $\chi = \vec{S}_i \bullet \vec{S}_j \times \vec{S}_k$ (Fig. 1a) accumulate a chirality-induced Berry phase that also contributes to $V_{Hall}$ [9,11-13]. Theoretically, the effect of the spin texture on the current can reach the equivalent of a local magnetic field of $H \sim 10^{4-5}$T, which is at least two orders of magnitude larger than the highest field currently generated in high magnetic field laboratories. Furthermore, if this local effective field does not average to zero over the sample, then the predicted Hall response can be as large as the integer quantum Hall effect in every atomic plane[14-17]. Because the local spin chirality leading to such a large effective magnetic field and associated Hall response can be tuned by applying a much smaller external field, a clean example of such an effect would open a new direction towards designing functional materials with exotic properties and states.

The highly-correlated Kondo lattice compound $UCu_5$ is a heavy-fermion metal in which the coupling between its U 5f-electron spins and band electrons leads to a large electronic specific heat coefficient of $\gamma \sim 300$mJ/molK$^2$, corresponding to an effective mass of its itinerant electrons nearly two orders of magnitude larger than that of a free electron[18]. Its magnetic U ions sit on a face-centred cubic (FCC) lattice and form a network of edge sharing tetrahedra--a geometry conducive to magnetic frustration[1] (Fig. 1b). Curie-Weiss fits to the magnetic susceptibility give a Weiss temperature of $\theta_W$=-284K which reflects the strong antiferromagnetic interactions between magnetic ions[19]. Yet, $UCu_5$ orders magnetically at temperatures $T<<\theta_W$, first at $T_N$=15.5K, and then at $T_2$=1.2K, due in part to magnetic frustration[1,2]. Powder neutron diffraction experiments indicate that the antiferromagnetic order in both phases consists of localized U 5f-electron spins pointing along the crystal unit cell's <111> directions with ordering wavevectors symmetry related to **q**=<1/2,1/2,1/2>[20-22], and the simplest magnetic order consistent with these data is the 1-q configuration shown in Fig. 1c with spins ferromagnetically aligned within (111) planes but antiferromagnetically aligned with neighbouring (111) planes. The magnitude of **q** does not change when cooling through $T_2$, despite clear signs of a phase transition in magnetization, resistivity, and specific heat data indicating that the transition is likely between single and multi-q states possessing the same |**q**|. In the absence of single crystals of $UCu_5$, it is impossible to distinguish a multi-domain 1-q configuration from more complex multi-q variants using powder neutron diffraction measurements. Nevertheless, nuclear magnetic resonance experiments have revealed significant differences in the distribution of the local exchange field acting on the nuclei, consistent with a transition between distinct states with different numbers of ordered Fourier components of magnetization[23]. Here we report a strikingly large geometrical Hall effect below $T_2$. On the basis of experimental results and theoretical analysis, we argue that the phase below $T_2$ possesses the 4-q non-coplanar antiferromagnetic order, illustrated in Fig. 1d, which causes a chirality-induced Berry phase in the itinerant electrons' wavefunctions and creates a large geometrical Hall effect that can be controlled by applying a relatively small magnetic field.

**Results**

**Magnetic Phase Diagram and Maximum in the Hall Resistivity**

The magnetic phase diagram for $UCu_5$ is shown in Fig. 2a, and we refer to the high and low temperature antiferromagnetic phases as M1 and M2, respectively. The phase diagram is constructed from $M(H)$ data discussed below as well as data shown in Supplementary Figures S1 through S7. M1 and M2 are phases possessing antiferromagnetic order, while PM is the paramagnetic phase. A Curie-Weiss fit to our susceptibility data is shown in Supplementary Figure S2 and gives $\theta_W$=-238(2)K, which is much greater than $T_N$ and agrees with previous results[19].

The Hall resistivity $\rho_{xy}(H)$ in both M1 and M2 is presented in Fig. 2b. Notably, in M2 $\rho_{xy}(H)$ has a pronounced peak near $H$~2-2.5T, and the corresponding Hall conductivity $\sigma_{xy}=-\rho_{xy}/(\rho_{xx}^2+\rho_{xy}^2)$ is extremely large, exceeding 20% of the $\nu=1$ quantum Hall effect per atomic layer at $H$=2.5T, which for free electrons with density of order one electrons per unit cell would require $H$~$10^3$T (Ref. 24). In contrast, $\rho_{xy}(H)$ in M1 is a simple monotonic function of magnetization, consistent with standard spin-orbit coupling dependent anomalous Hall effect mechanisms, such as skew scattering [9,10] (Supplementary Figure S7). The solid lines in Fig. 2b are data from a second separately prepared $UCu_5$ sample confirming the peak in $\rho_{xy}(H)$ is intrinsic to M2.

**Magnetization and Magnetoresistance**

To examine the possible mechanisms responsible for the observed maximum in the low-temperature Hall resistance, the field- and temperature-dependent Hall responses should be compared to $M(H)$ and $\rho_{xx}(H)$ for temperatures and fields spanning M1 and M2. Figure 3a is an image plot of $M(H)$ in which different colours correspond to $dM/dH$ and highlight the M1-M2 phase boundary. In M1, $M(H)$ is linear for fields up to the M1-M2 phase boundary where there is a step increase. After the step, $M(H)$ continues linearly in M2 but with a larger slope, and $M(H)$ is linear in M2 for all temperatures and fields measured. Figure 3b shows that at $T=1.4$ and 1.8K $\rho_{xx}(H)$ exhibits a positive magnetoresistance in M1, with hysteresis occurring at the M1-M2 phase boundary. Once a high enough field is reached to cross into M2, $\rho_{xx}(H)$ has negative magnetoresistance in agreement with $T=0.05$-0.2 K data. $T=1$K data show large hysteresis near the M1-M2 boundary, which is consistent with the previously reported sharp temperature-hysteretic peak at $T_2$ in specific heat indicative of a first order transition[18]. $\rho_{xx}(H)$ in M2 shows a difference between virgin and field cycled curves but no other hysteresis.

**Ruling Out Conventional Anomalous Hall Effect Mechanisms**

The highly unconventional peak in $\rho_{xy}(H)$ in M2 contrasts with the featureless behaviour of both $\rho_{xx}(H)$ and $M(H)$. At $T<0.5$K, Fig. 3b shows that $\rho_{xx}(H)$ is almost constant up to $H=3$T, even as a peak in $\rho_{xy}(H)$ is reached at $H\sim2$-2.5T. Similarly, in M2 $M(H)$ is linear in $H$ and does not show any anomalies around $H\sim2$-2.5T. Therefore, in contrast to $\rho_{xy}(H)$ in M1, the unusual behaviour of $\rho_{xy}(H)$ in M2 is not dominated by the standard anomalous Hall effect due to the spin-orbit interaction. The differences in behaviour between $\rho_{xy}(H)$ in the M1 and M2 phases is particularly clear in the derivative $d\sigma_{xy}/dM$ (Supplementary Figure S8a). In the low field region of M2 the behaviour of $\sigma_{xy}$ is highly unconventional with $d\sigma_{xy}/dM$ changing sign with increasing $H$, and Supplementary Figure S8b shows that the maximum in $\rho_{xy}(H)$ cannot be fit assuming skew scattering alone. This is the regime in which a geometrical Hall effect due to the acquired chirality-induced Berry phase dominates.

**Discussion**

To demonstrate how a chirality-induced geometrical Hall effect arises due to the non-coplanar 4-q magnetic order, we analyze the properties of $UCu_5$ within the Kondo lattice model (KLM) for an FCC lattice:

$$H_{KLM} = -t \sum_{\langle i,j \rangle \sigma} (c_{i\sigma}^\dagger c_{j\sigma} + c_{j\sigma}^\dagger c_{i\sigma}) - J_K \sum_{i\alpha\beta\delta} (c_{i\alpha}^\dagger \sigma_{\alpha\beta}^\delta c_{j\beta}) S_i^\delta \quad (1)$$

where $\sigma^\delta$ are the Pauli matrices, $c$ is the annihilation operator for the itinerant electrons, $\langle i, j \rangle$ indicates nearest-neighbour U sites, $t$ is the hopping matrix element for the current of itinerant electrons, and $S^\delta$ is the component of the U spin in the $\delta$ direction. The dynamical Kondo coupling $J_K$ gives the coupling strength between itinerant electrons and U spins and is the origin of the large effective mass[18]. Here, we will concentrate on the magnetically ordered phases and assume that local moments are classical. Using a variational calculation[24] we find that depending on $J_K$ and the carrier density of the current, $n$, both the 1-q and 4-q states, among others, can be stabilized at $T=0$. A typical parameter set to stabilize 4-q order is $J_K/t=3.5$ with $n=0.3$ electrons per site. It is possible to study the thermodynamic phase transitions of the KLM directly[24], but it is very computationally-intensive. Therefore, for finite temperature calculations we use a related classical spin model which gives the same result as the KLM at $T=0$,

$$H_{CS} = J_1 \sum_{\langle i,j \rangle} S_i \cdot S_j + J_2 \sum_{\langle\langle i,j \rangle\rangle} S_i \cdot S_j + K \sum_{\langle i,j \rangle} (S_i \cdot S_j)^2 - \sum_i H \cdot S_i \quad (2)$$

Here, $S$ is the spin of Uranium, $J_1$ and $J_2$ are the antiferromagnetic exchange coupling strengths between nearest and next nearest-neighbours, respectively, $K$ is the nearest-neighbour bi-quadratic exchange strength, and the final term allows coupling to an applied magnetic field. For $2J_2>J_1$ and $K=0$ the model yields a continuum of ground states due to magnetic frustration. Single spin flip Monte-Carlo simulations using $J_1=1.5$, $J_2=1$, and $K=0.01$, which give the experimentally observed ratio of $T_N/T_2$, yield the $H=0$ specific heat curve shown in Fig. 4a. The simulations reproduce the two phase transitions, with the lower temperature phase possessing the 4-q order and the higher temperature phase possessing the 1-q order. (Within the framework of these simulations, the transitions at $T_2$ and $T_N$ appear to be first and second order, respectively, which is consistent with experiments. The order of the transitions, however, has not yet been confirmed with complete certainty due to computational limitations.) The transition between M2 and M1 is driven by thermal fluctuations favouring the 1-q state, because the order-by-disorder mechanism stabilizes coplanar magnetic order at finite temperature for a frustrated FCC lattice[25,26]. (We note that the analysis of the NMR data in Ref. 23 gives M1 as the 4-q state and M2 as the 1-q state. However, the 4-q state is expected to be more energetically favorable than the 1-q state in Kondo-lattice systems. Indeed, a highly frustrated insulator $Gd_2Ti_2O_7$ has been shown to have a 4-q ground state and a 1-q state at higher temperatures [26]. This trend also is consistent with general theoretical arguments that thermal fluctuations tend to stabilize the simpler magnetic configuration[25]. Moreover, $\rho_{xx}(T)$ data for $Lu_{0.02}U_{0.98}Cu_5$ shown in Supplementary Figure S9 show an increase in $T_2$ with non-magnetic dilution which is also consistent with order-by-disorder.) Figures 4b and 4c show snapshots taken during the simulations of the spin configurations in the 4-q and 1-q phases, respectively. In the KLM we have treated the spins classically and have not assumed any spin-orbit coupling which is the key ingredient in most anomalous Hall effect theories and has been shown to be important in explaining the Hall effect in $Pr_2Ir_2O_7$ and $Nd_2Mo_2O_7$[27].

Unlike spins, the scalar spin chirality does not couple directly to a magnetic field and is invariant to the rigid rotation of all the spins that can be induced by a rotation of an external magnetic field. Therefore, it may appear that the Hall effect attributable to the scalar spin chirality must average to zero for polycrystalline samples. The presence of itinerant electrons prevents this because electrons moving in a chiral texture acquire *orbital* magnetization[28] that directly couples to the magnetic field. For the experimental values of $\sigma_{xy}$ calculated from data in Fig. 2b, the orbital magnetization per unit cell is estimated to be $\sim 10^{-2}\mu_B$; therefore, at $T\sim 1K$, a magnetic field of only 1 Tesla should be sufficient to align even small domains containing 100 unit cells. Thus, the acquired orbital magnetization can produce an experimental signature of a large geometrical Hall effect, even in polycrystalline samples.

Given this coupling between the orbital magnetization and magnetic field, we calculate $\sigma_{xy}(H)$ at $T=0$ by first choosing among the multiple energetically equivalent ground states of equation (2) with different values of the scalar spin chirality $\chi_{111}$ projected on to the [111] direction, which for the purpose of our calculations we assume to be parallel to **H**. We then select the state with the largest $\chi_{111}$ to give us an initial spin configuration as this state will be favoured by the coupling between the orbital magnetization and **H**. Figure 5a shows $\chi_{111}(\mathbf{H})$ and examples of the spin configurations for different **H**. Next, we substitute the determined spin configurations into equation (1) and use the Kubo formula[29] to calculate the chirality-induced Berry phase contribution to $\sigma_{xy}(H)$. Figures 5b and 5c show the measured and calculated $\sigma_{xy}(H)$ curves, respectively. Remarkably, both curves show a local minimum as a function of magnetic field and have comparable magnitudes, despite the simplistic way in which we have chosen the Hamiltonian parameters. Though qualitative in nature, these calculations demonstrate that the magnetic field distortion of the 4-q magnetic order can induce an extremely strong response in the Hall conductivity, comparable to experimental results and illustrate the principle of generating large geometric Hall responses with a field due to a non-coplanar spin texture without assuming any material-specific spin-orbit interaction.

These results strongly suggest that the peak in $\rho_{xy}(H)$ in the M2 phase is due to a chirality-induced Berry phase caused by the frustration-created spin texture. Although previous works on $Nd_2Mo_2O_7$ and $Pr_2Ir_2O_7$ have argued for observation of such a geometrical Hall effect [12,13], it has been contested that in these cases the Hall effect primarily stems from the magnetic Mo or Ir ions and is due to the d-orbital angular momentum induced by the spin-orbit interaction rather than being induced by the non-coplanar spin textures of the Nd or Pr[27]. In our case, Cu is not magnetic and the chirality-induced Berry phase is due to magnetic exchange between f and d electrons. Also, the Berry phase due to spin-orbital effects in the electronic band structure should be insensitive to small amounts of disorder. As we show in Supplementary Figures S9 and S10 and discuss in the Supplementary Methods section of the Supplementary Information, Hall measurements on $UCu_{5-x}M_x$, with $M$=Ag or Au and $x$=0.01, 0.03 (i.e. 0.2 and 0.6% substitution) and $U_{1-x}Lu_xCu_5$, $x$=0.02 show that the peak in $\rho_{xy}(H)$ is suppressed by small, nominally isoelectronic substitutions. These experiments rule against a notable contribution to the Berry phase from the spin-orbit induced momentum-space monopoles[9] and demonstrate that the 4-q order in $UCu_5$ is necessary for the observed geometrical Hall effect. Finally, a chirality-induced Berry phase effect also has been proposed to explain $\sigma_{xy}$ in MnSi[30,31] and MnGe[32]; however, in these cases the induced Hall effect is orders of magnitude weaker than in $UCu_5$ due to the very smooth nature of the Skyrmion[33] textures.

We have shown that a tunable, unusually large Hall conductivity, equivalent to the presence of an effective magnetic field of $\sim 10^3$T, arises in $UCu_5$ and that the chirality-induced Berry phase due to frustration-induced non-coplanar 4-q antiferromagnetic order creates this large geometrical Hall effect. These results provide an example of the possible exotic transport properties resulting from magnetic frustration in highly-correlated metals, while illustrating the opportunities afforded by pursuing studies on other frustrated metals. In particular, control of the geometrical Hall effect by a small magnetic field shows that frustration can be an important tuning parameter, allowing for the tailoring of specific properties and states of correlated matter.

**Methods**

**Sample Synthesis**

Polycrystalline samples of $UCu_{5.1}$ and $UCu_{5.05}$ were prepared by arc melting on a water-cooled Cu hearth under an Ar atmosphere, followed by annealing under vacuum ($10^{-4}$ Torr) at 800 or 850°C for 1-3 weeks. Samples were determined to be single phase by x-ray diffraction, and we confirmed that both $UCu_{5.1}$ and $UCu_{5.05}$ share the same field-temperature phase diagram. Hence, for simplicity we refer to both samples as $UCu_5$. The $UCu_{5-x}M_x$, $M$=Ag, Au, $x$=0.01, 0.03, an $U_{0.98}Lu_{0.02}Cu_5$ samples were made in a similar fashion.

**Magnetization Measurements**

Magnetization measurements were made in a Quantum Design Superconducting Quantum Interference Device magnetometer, in a dilution refrigerator utilizing either a capacitance-based Faraday or torque magnetometer, and in pulsed magnetic fields using a compensated coil-extraction magnetometer.

**Resistivity Measurements**

Four-wire resistivity measurements of the longitudinal resistivity and Hall resistivity were performed on thin rectangular samples using a Linear Research LR-700 ac resistance bridge, and the samples were cooled and exposed to a magnetic field in a Quantum Design Physical Property Measurement System or a dilution refrigerator with an incorporated magnet. Pt leads were attached to the samples by either spot welding or by applying silver paint. Longitudinal contributions to the Hall resistivity were cancelled by making measurements in both positive and negative fields. High field measurements were made at the National High Magnetic Field Laboratories in Tallahassee and Los Alamos.

**Supplementary Information** is linked to the online version of the paper at www.nature.com/nature.

**Acknowledgments:**

We thank C. Pfleiderer, C.D. Batista, V.S. Zapf, S. Brown, G. Koutroulakis, and M.F. Hundley for discussions and assistance. Work at Los Alamos was performed under the auspices of the U.S. Department of Energy and supported by the Laboratory Directed Research and Development program. Z.F. acknowledges support from the National Science Foundation NSF-DMR-0801253. B.G.U. acknowledges support from the G.T. Seaborg Institute for Transactinium Science.

**Author Contributions:**

P.H.T., M.A.T., Z.F., and E.D.B. made the samples. C.F.M., R.O., R.M., Z.F., O.A-Y., R.D.M, F.R., and B.G.U. made resistivity and Hall measurements. J.D.T., C.F.M., R.O., R.M., O.A-Y., and R.D.M made magnetization measurements. O.A-Y and R.D.M performed the high magnetic field measurements. Y.K. and I.M. developed the theory and performed calculations. J.D.T. suggested the course of study. B.G.U. performed analysis and wrote the manuscript with editorial input from all of the authors.

The authors declare no competing financial interests.

**Figure Captions**

Fig. 1 Affect of spin chirality on an itinerant electron and UCu$_5$'s antiferromagnetic order. (a) A diagram illustrating the geometrical (Berry) phase $\varphi$ gained by an itinerant electron's wavefunction traversing the three non-coplanar spins S$_i$, S$_j$, and S$_k$ possessing finite scalar spin chirality $\chi$. $\Psi_{e^-}$ is the incident wavefunction of the electron. (b) Illustration of the U sublattice of edge-sharing tetrahedra. (c) Illustration of the 1-q type magnetic order where **q** = <1/2,1/2,1,2>. (d) Illustration of the 4-q type of magnetic order where **q** = <1/2,1/2,1,2>. For the 1-q order, spins (red arrows) lie along [111], are ferromagnetically aligned within (111) planes, and antiferromagnetically aligned between them; whereas, the 4-q magnetic order is non-coplanar, with spins pointing along various <111> directions with 4 propagation vectors: **q** = [1/2,1/2,1,2], [-1/2,1/2,1,2], [1/2,-1/2,1,2], and [1/2,1/2,-1,2]. For the 4-q order, an electron traversing the non-coplanar spins acquires a chirality-induced Berry phase due to the mechanism shown in (a). Illustrations in (b),(c), and (d) were made using VESTA[36].

Fig. 2 Magnetic phase diagram and the Hall resistivity. (a) The phase diagram for UCu$_5$. M1 and M2 are phases possessing the 1-q and 4-q antiferromagnetic order, respectively, while PM is the paramagnetic phase. The frustration-induced phase boundary between M1 and M2 at low fields agrees with earlier results[34]; however, we do not see the previously reported multiple phase transitions at higher fields, which is likely due to subtle differences between our samples and those in Ref. 34 (see also Ref. 35 and Supplementary Figure S4). These differences are inconsequential to our conclusions. (b) The Hall resistivity $\rho_{xy}$ for temperatures and fields spanning M1 and M2. Solid blue

circles are $T=0.05$K data, open red circles are $T=0.2$K data, solid green triangles are $T=0.5$K data, open black triangles are $T=1.0$K data, solid purple squares are $T=1.4$K data, and open wine squares are $T=1.8$K data. In M1 ($T=1.8$ and 1.4K), $\rho_{xy}(H)$ is a simple function of $M$ and $\rho_{xx}$ and can be described by the usual anomalous Hall effect mechanisms discussed in the Supplementary Methods section of the Supplementary Information, but in M2 ($T=0.05 - 1$K) $\rho_{xy}(H)$ exhibits a peak near $H\sim2$-2.5T consistent with a geometrical Hall effect. Solid lines are $T=0.65$K (blue) and $T=0.85$K (black) data from a second separately prepared sample confirming the peak is intrinsic to M2.

Fig. 3 Magnetization and longitudinal resistivity for temperatures and fields spanning M1 and M2. (a) Image plot of the field dependence of the magnetization $M$ for various temperatures. Different colours indicate the slope d$M$/d$H$. $M(H)$ is linear in both M1 and M2, with a change of slope occurring at the phase boundary. (b) The field dependence of the longitudinal resistivity $\rho_{xx}$ for various temperatures. Solid blue circles are $T=0.05$K data, open red circles are $T=0.2$K data, solid green triangles are $T=0.5$K data, open black triangles are $T=1.0$K data, solid purple squares are $T=1.4$K data, and open wine squares are $T=1.8$K data. For $T<1$K, $\rho_{xx}(H)$ is nearly constant up to $H\sim2.5$T, but for $T=1.8$ and 1.4K, hysteresis occurs at $H\sim7$ and 4T, respectively, as the M1-M2 boundary is crossed. At $T=1$K hysteresis occurs over most of the field range, consistent with the steep phase boundary.

Fig. 4 Monte-Carlo simulation results reproducing the experimentally observed magnetic phase transitions. (a) The calculated heat capacity $C$ versus the reduced temperature

$T/J_2$ using $J_1=1.5$, $J_2=1$, and $K=0.01$. The different curves correspond to different lattice sizes ($L=4$: red squares; $L=6$: blue circles; $L=8$: black triangles). (b) A snapshot of the spin structure taken during the Monte-Carlo simulations showing that 4-q antiferromagnetic order exists below the lower temperature phase transition. (c) A snapshot of the spin structure taken during the Monte-Carlo simulations showing that 1-q antiferromagnetic order exists between the lower and higher temperature phase transitions.

Fig. 5 Field-controlled geometrical Hall effect due to the frustration-induced 4-q magnetic order. (a) The scalar spin chirality $\chi_{111}$ at $T=0$ as a function of the external magnetic field, $H$. Examples of spin configurations obtained by mean field analysis of the Hamiltonian (2) corresponding to different values of $\chi_{111}$ are also shown. (b) The calculated $T=0$ Hall conductivity $\sigma_{xy}$ due to the geometrical Hall effect as a function of magnetic field $H$ in the frustration-induced 4-q magnetically ordered state ($J_1=1.5$, $J_2=1$, $K=0.01$, $J_K/t=3.5$, $n=0.3$). (c) The experimentally observed Hall conductivity as a function of $H$. Solid blue circles are $T=0.05$K data, open red circles are $T=0.2$K data, and solid green triangles are $T=0.5$K data. In the shaded region of (b), the calculation reproduces the minimum in the experimental data shown in (c) along with the rise of $|\sigma_{xy}|$ at higher fields. The calculated values of $\sigma_{xy}(H)$ arise solely from the acquired chirality-induced Berry phase and do not include material-specific spin-orbit effects. Similar calculations assuming 1-q magnetic order do not yield a finite chirality-induced Berry phase contribution to $\sigma_{xy}(H)$.

(a)
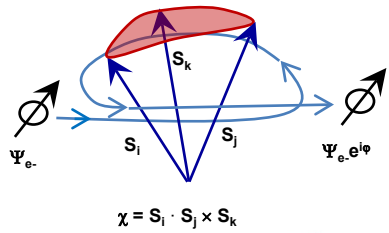

$\chi = S_i \cdot S_j \times S_k$

(b) 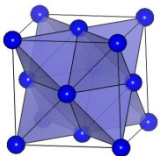
U only

(c) 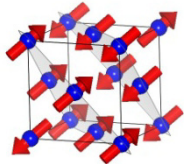
1-q

(d) 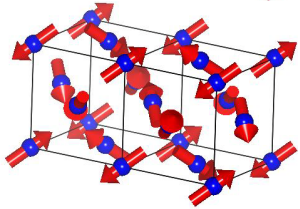
4-q

(a)

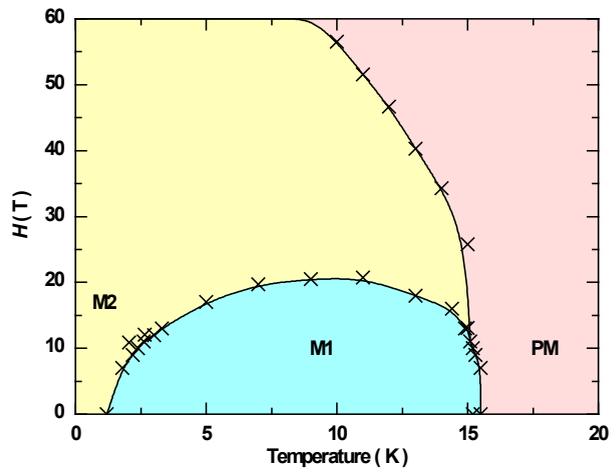

(b)

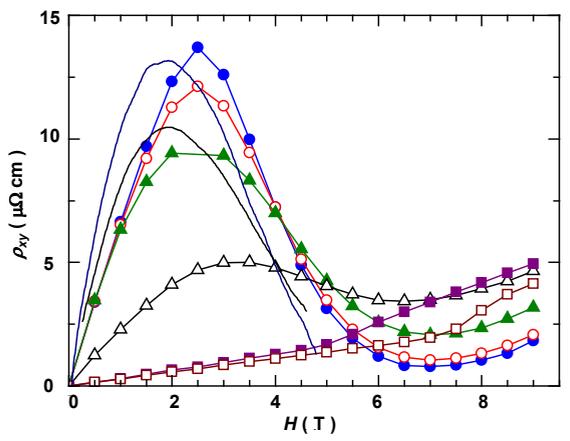

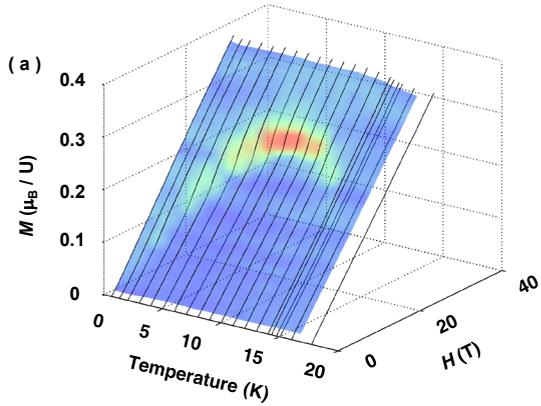

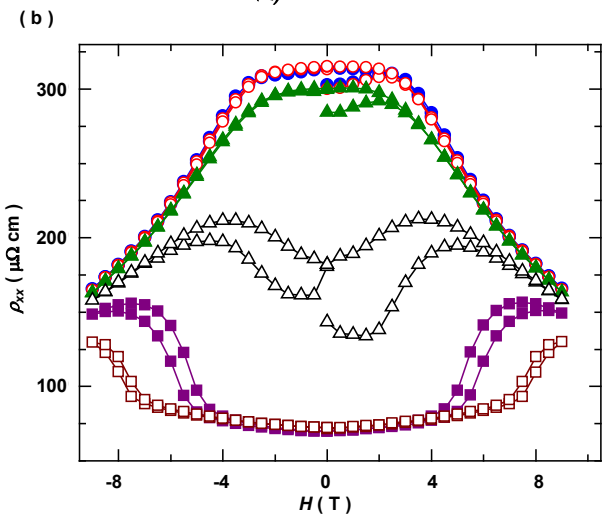

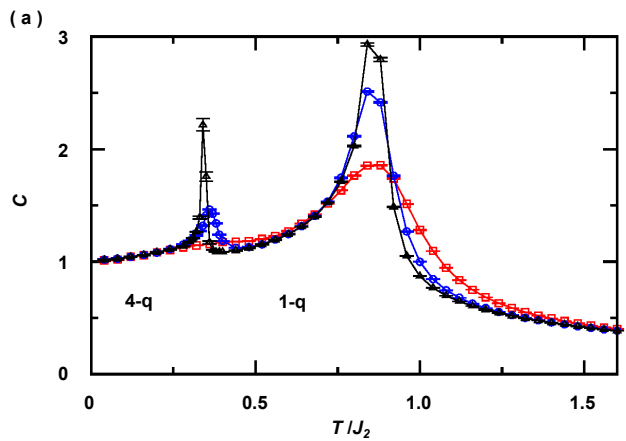
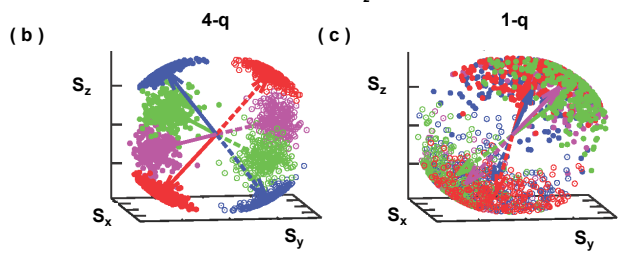

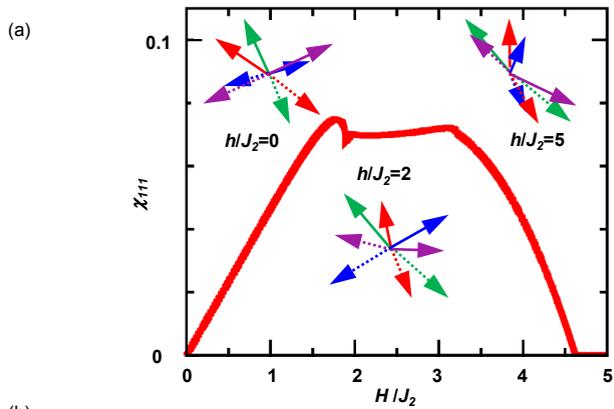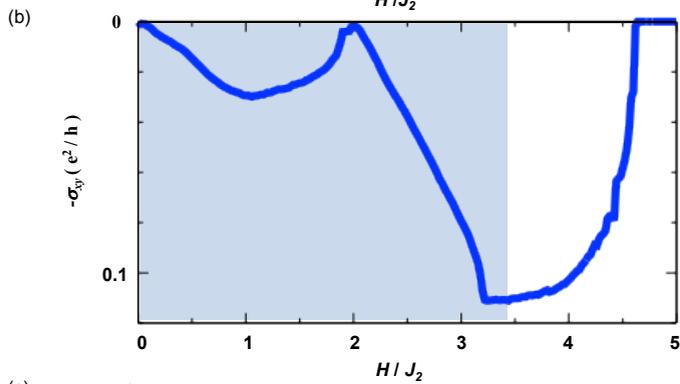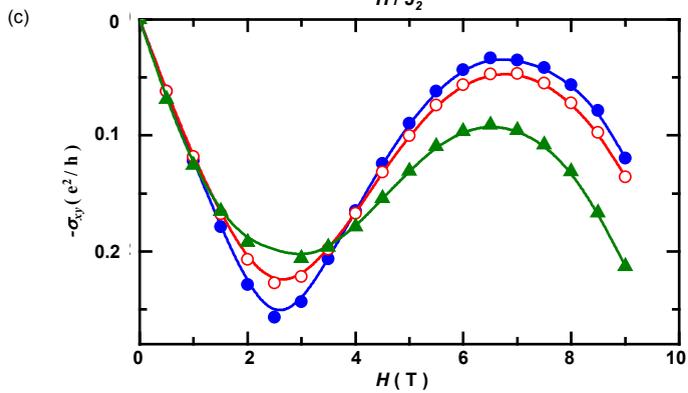

**Supplementary Information**

**Controllable chirality-induced geometrical Hall effect in a frustrated highly-correlated metal**

B. G. Ueland, C. F. Miclea, Yasuyuki Kato, O. Ayala–Valenzuela, R. D. McDonald, R. Okazaki, P. H. Tobash, M. A. Torrez, F. Ronning, R. Movshovich, Z. Fisk, E. D. Bauer, Ivar Martin, and J. D. Thompson

**Supplementary Figures**

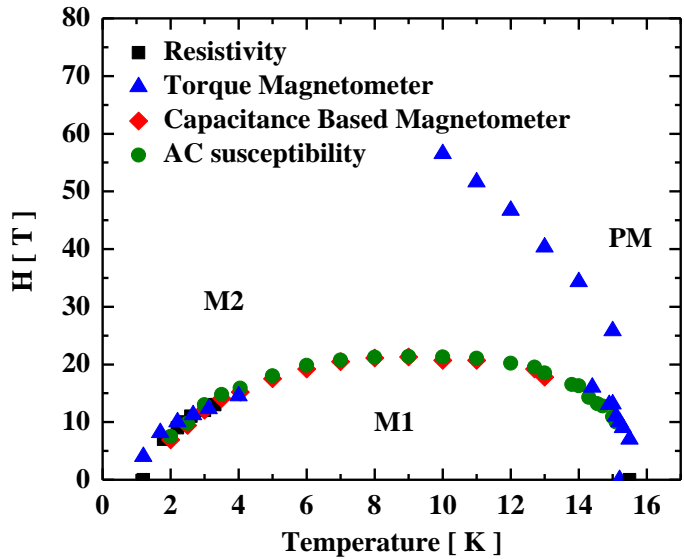

**Supplementary Figure S1. Phase diagram of UCu$_5$ determined from magnetization, ac susceptibility, and resistivity measurements.** Magnetization measurements were performed using a capacitance based Faraday magnetometer or a torque magnetometer. Points on the phase diagram were determined from both isothermal data taken while changing the field and from constant field data taken while sweeping the temperature. M1 and M2 refer to the high and low temperature antiferromagnetic phases, respectively, and PM denotes the high temperature paramagnetic phase. Note that different measurements give consistent results for the phase boundaries.

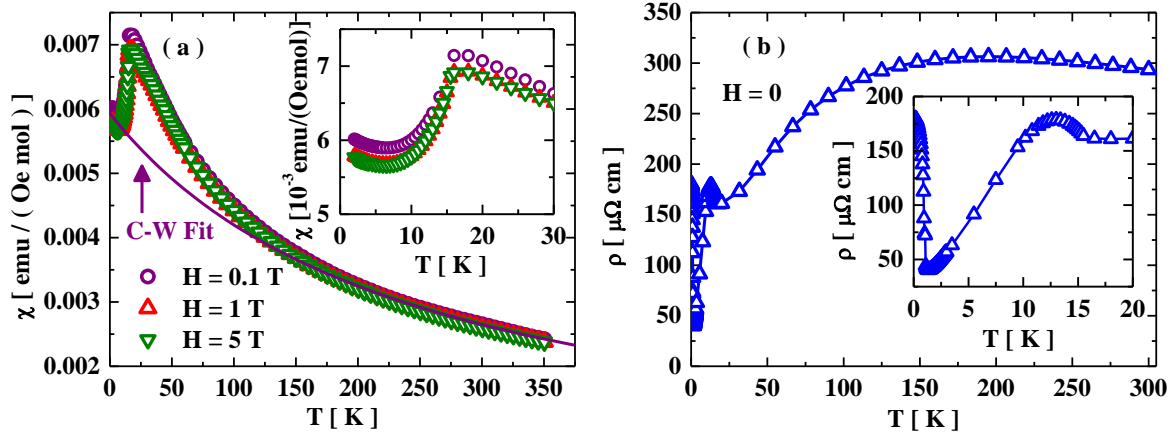

**Supplementary Figure S2. Susceptibility $\chi$ and longitudinal resistivity $\rho_{xx}$ of UCu$_5$.** (a) $\chi(T)$ of UCu$_5$ measured in a Quantum Design SQUID magnetometer at various fields. The solid line shows a Curie-Weiss fit to the $H$=0.1T data over $T$=350-250K. The Weiss temperature determined from the fit is $\theta_W$=-238(2)K and the effective moment is $p$=3.39(1)$\mu_B$/U. The inset shows the drop in susceptibility associated with the transition from the paramagnetic to M1 antiferromagnetic phase at $T_N$=15K. (b) $\rho_{xx}(T)$ of UCu$_5$ at $H$=0. Cooling from $T$=300K, $\rho_{xx}(T)$ first has a maximum at $T$~170K and then decreases until reaching $T_N$. The inset shows a peak occurring just below $T_N$ that reflects partial gapping of the Fermi surface as the sample is cooled into the M1 phase. At $T_2$, $\rho_{xx}(T)$ quickly rises with decreasing temperature. $\rho_{xx}(T)$ at $H$=0 is lower in the M1 phase than in either the paramagnetic or M2 phases (Supplementary Figure S6).

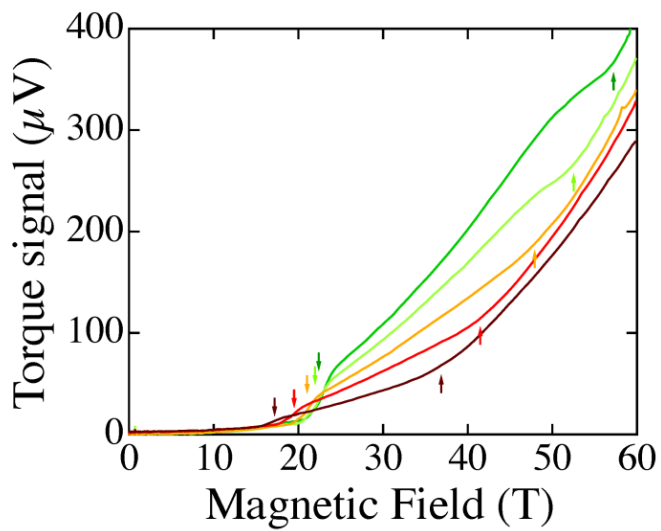

**Supplementary Figure S3. Magnetic torque data used to determine the phase transitions between *T*=10-14K.** The green curve is the *T*=14K data while the brown curve is the *T*=10K data. Points for the phase diagram are taken at fields where changes in slope occur and are indicated by arrows.

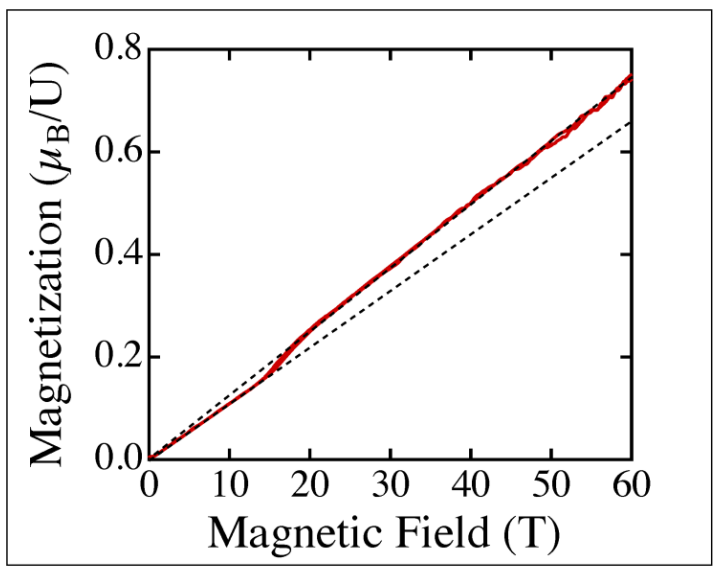

**Supplementary Figure S4. Magnetization *M* versus field *H* at *T*=4K.** The change in slope in the data indicates the transition from M1 to M2. There is no evidence for any other phase transitions up to *H*=60T.

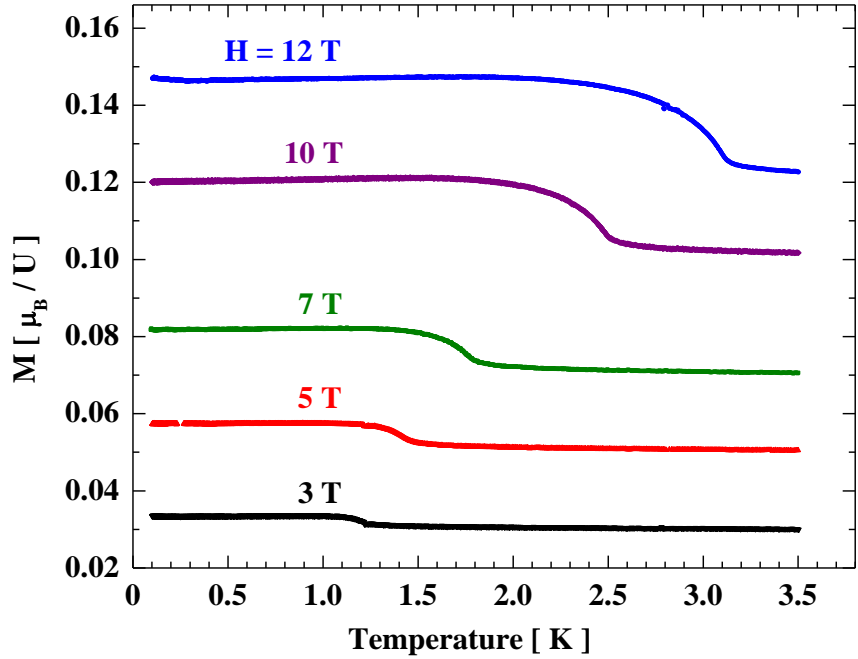

**Supplementary Figure S5. Low temperature magnetization *M* of UCu$_5$.** Measurements were made in a dilution refrigerator using a capacitance-based Faraday magnetometer, and data are calibrated to *T*=2K data taken in a Quantum Design SQUID Magnetometer. A sharp step in *M* is seen at the phase transition between M1 and M2, and the temperature of the phase transition increases with increasing field.

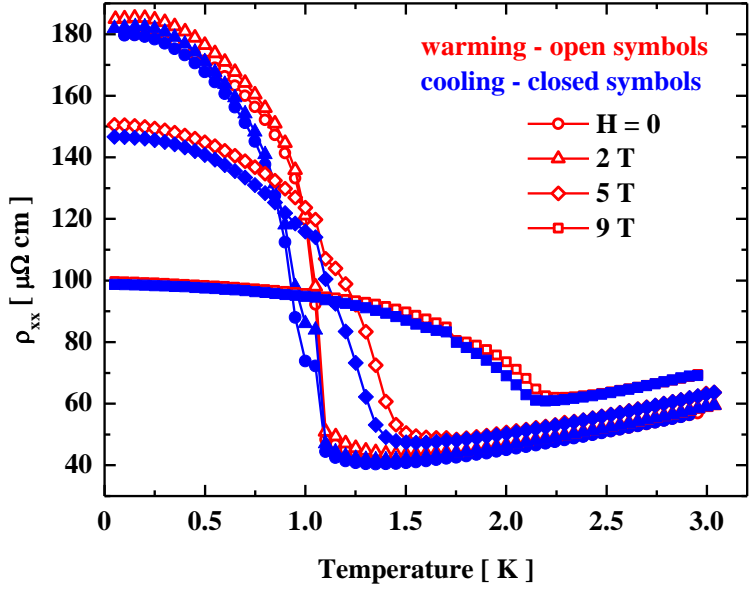

**Supplementary Figure S6. Low temperature longitudinal resistivity $\rho_{xx}$ at various magnetic fields $H$.** Data were taken during warming and cooling with thermal hysteresis occurring at the phase boundary between the M1 and M*2* antiferromagnetic phases. $\rho_{xx}$ in M*2* increases substantially as expected for 4-q type magnetic order creating additional gapping of the Fermi surface. The temperature of the phase transition increases with increasing field. In M2 $\rho_{xx}$(T) decreases with increasing field as the chiral spin texture rotates along the field direction.

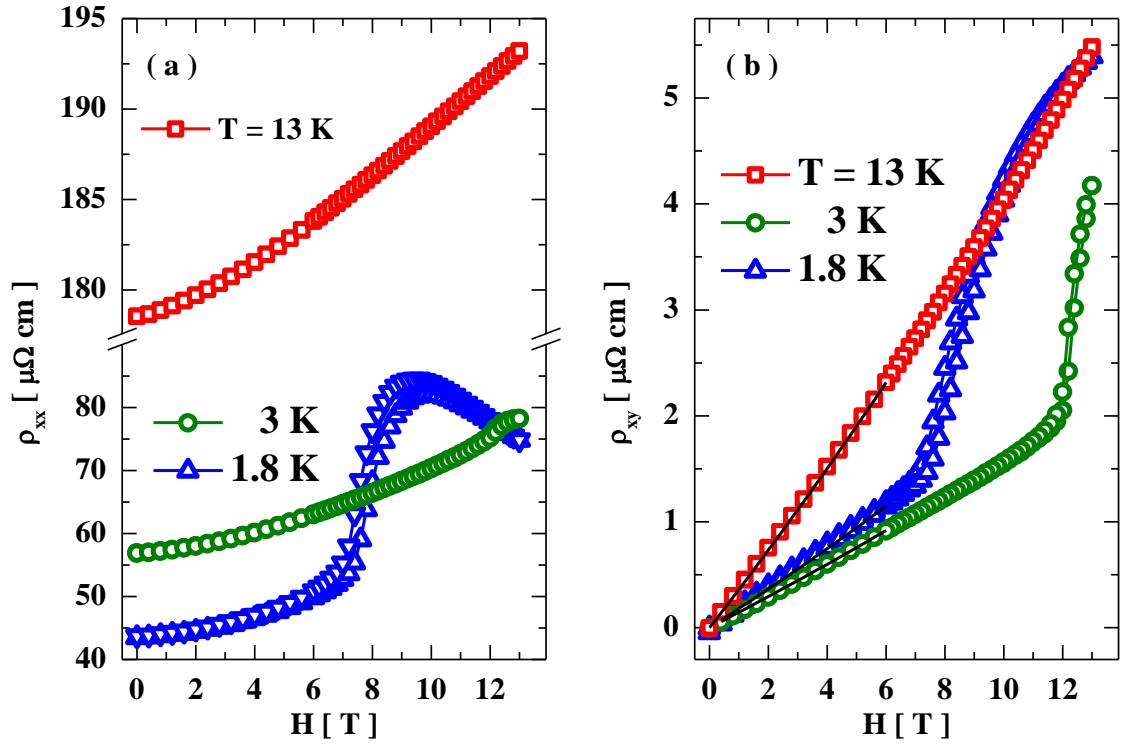

**Supplementary Figure S7. Magnetic field $H$ dependence of the longitudinal $\rho_{xx}$ and transverse $\rho_{xy}$ resistivity for temperatures spanning M1.** (a) $\rho_{xx}(H)$ at various temperatures. $\rho_{xx}(H)$ increases with increasing field in the M1 phase. The $T$=1.8K data show a step and associated hysteresis at the phase boundary between M1 and M2 (a small step also occurs in the $T$=3K data). $\rho_{xx}(H)$ decreases with increasing field after entering M2. (b) The Hall resistivity $\rho_{xy}(H)$ at various temperatures. The solid lines between $H$=0-6T are fits to $\rho_{xy}$=R$_0$H + R$_A$M$\rho_{xx}$, where $M$ is the magnetization and R$_0$ and R$_A$ are constants multiplying the normal and the skew scattering Hall terms, respectively. The fits show that in M1 $\rho_{xy}$ is well described by the usual ordinary and skew-scattering Hall terms, and is not due to a Berry phase induced geometrical Hall effect. The steep rise and hysteresis in $\rho_{xy}(H)$ at high $H$ in the $T$=1.8 and 3K data are due to the phase transition between M1 and M2.

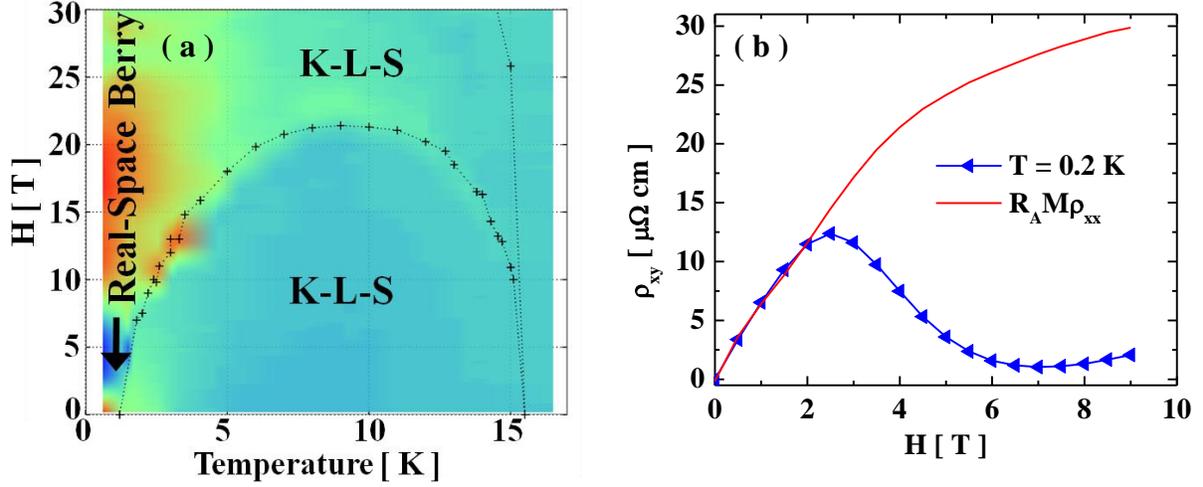

**Supplementary Figure S8. Regions dominated by either the geometrical Hall effect or skew scattering.** (a) The phase diagram coloured to show changes in the derivative of the Hall conductivity with respect to the magnetization, $d\sigma_{xy}/dM$. Regions labelled K-L-S correspond to $\sigma_{xy} \sim M$, while $d\sigma_{xy}/dM$ changes sign in the low temperature, low field region of M2 where the geometrical Hall effect occurs. (b) An attempt to fit the maximum in the transverse resistivity $\rho_{xy}(H)$ in M2 to the skew-scattering Hall term $\rho_{xy} \sim R_A M \rho_{xx}$. $M$ is the magnetization in $\mu_B$/U, $\rho_{xx}$ is the longitudinal resistivity, and $R_A$ is a constant. $M(H)$ increases linearly with increasing $H$, while $\rho_{xx}(H)$ is constant until $|H| \sim 2$-$2.5$ T and then starts decreasing with increasing $|H|$ (Fig. 3). The maximum in $\rho_{xy}(H)$ cannot be reproduced assuming skew scattering alone.

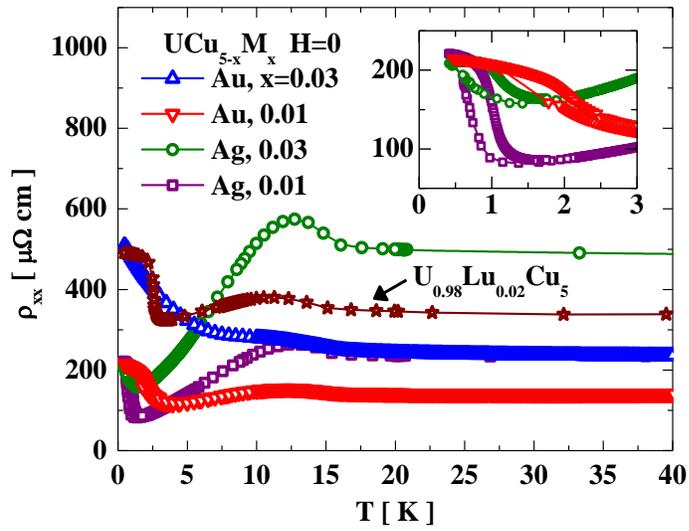

**Supplementary Figure S9. Zero field longitudinal resistivity $\rho_{xx}$ versus temperature $T$ data for $UCu_{5-x}M_x$ and $U_{0.98}Lu_{0.02}Cu_5$, $M$=Ag, Au, $x$=0.01, 0.03.** The inset shows the hysteresis present in the Ag and $x$=0.01 Au substituted samples. Similar hysteresis occurs in the Lu $x$=0.02 sample.

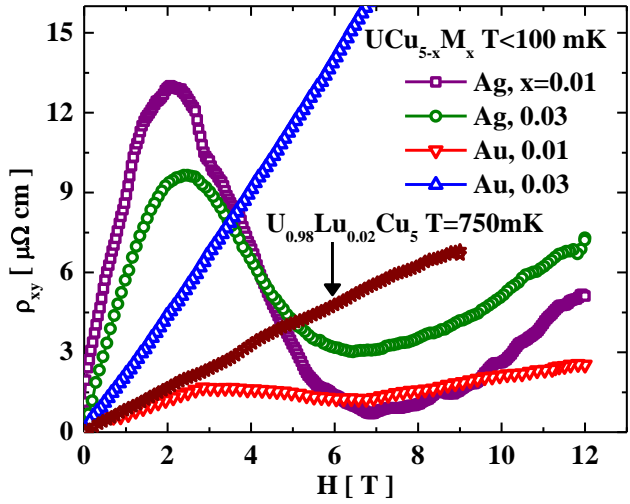

**Supplementary Figure S10. Hall resistivity $\rho_{xy}$ versus field $H$ for $UCu_{5-x}M_x$ and $U_{0.98}Lu_{0.02}Cu_5$. $M$=Ag, Au, $x$=0.01, 0.03.** The maximum in $\rho_{xy}(H)$ decreases with increasing Ag doping, is small but finite for Au $x$=0.01, and does not exist for Au $x$=0.03 and $U_{0.98}Lu_{0.02}Cu_5$.

**Supplementary Methods**

**Contributions to the Hall Effect**

The Hall effect[37] occurs when a magnetic field is applied perpendicular to an electric current. Due to the Lorentz force:

$$\vec{F} = q\vec{v} \times \mu_0 \vec{H}, \tag{S1}$$

the current is deflected by the field and a Hall voltage is generated transverse to the current. In magnetic materials, an anomalous Hall effect can occur without applying an external magnetic field due to the interaction between itinerant charges and spin degrees of freedom. This is illustrated in the simple case of a metallic ferromagnet where the magnet's net uniform magnetization in combination with spin-orbit interaction acts as an effective magnetic field creating the Hall voltage. Magnetically polarized electrons acquire an anomalous velocity due to the spin-orbit interaction and develop a Hall resistance that is proportional to the material's magnetization $M$ and the square of its longitudinal resistivity $\rho_{xx}$ – the so called Karplus and Luttinger[38] (K-L) contribution.

A criticism of the Karplus-Luttinger theory is that it does not take into account magnetic scattering of electrons off of magnetic impurities. An extension of this theory that includes the spin-orbit interaction due to asymmetric scattering off of magnetic impurities[39] leads to a term in the anomalous Hall effect that is proportional to $M$ and $\rho_{xx}$. This contribution to the anomalous Hall effect is referred to as skew scattering, and skew scattering contributions typically are observed, for example, in non-magnetic metals containing rare-earth impurities as well as in non-frustrated heavy-fermion materials[40].

**Evidence Against Spin-Orbit Induced Momentum-Space Magnetic Monopoles in the Band Structure Significantly Contributing to the Peak in $\rho_{xy}(H)$**

Comparing the resistivity $\rho_{xx}$ curves in Supplementary Figure S9 with those in Supplementary Figures S2b and S6 shows that a signature of entry into the M2 (4-q) magnetic phase persists with Ag doping (i.e. a step-like increase in $\rho_{xx}(T)$ upon cooling into the M2 phase), with a corresponding peak in the Hall resistivity $\rho_{xy}$ (Supplementary Figure S10) similar to data for $UCu_5$. The signature of the transition into M2 is broadened for $x=0.01$ Au and completely absent for $x=0.03$ Au, and there is a much smaller ($x=0.01$) or no ($x=0.03$) anomalous peak in $\rho_{xy}(H)$. $\rho_{xx}(T)$ data for the $x=0.02$ Lu substituted sample shown in Supplementary Figure S9 also show the signatures of entry into the M1 and M2 phases upon cooling but no anomalous peak in the Hall resistivity in the M2 phase (Supplementary Figure S10). Furthermore, $\rho_{xx}(T)$ data for both Ag samples, the $x=0.01$ Au sample, and the $x=0.02$ Lu sample show hysteresis at the M1-M2 phase boundary (inset to Supplementary Figure S9), while data for the $x=0.03$ Au sample do not show hysteresis at low $T$. $\rho_{xx}(H)$ data for both Ag samples, $x=0.01$ Au, and $x=0.02$ Lu are also similar to data for $UCu_5$ in Fig. 3b and show a crossover from positive to negative magnetoresistance upon cooling from M1 into M2, while the magnetoresistance for the $x=0.03$ Au sample is negative throughout the temperature range studied. Indeed, for $x=0.03$ Au doping $\rho_{xx}$ is much larger than that for $UCu_5$ and the other samples at the lowest temperatures measured and $\rho_{xy}$ Au is approximately linear in field. For $UCu_5$, skew scattering contributes to the Hall effect, but it is subdominant to the chirality-induced Hall effect due to the 4-q structure in the M2 phase. When 4-q order is absent, as in $x=0.03$ Au, skew scattering, together with the much larger magnitude of $\rho_{xx}$ in $x=0.03$ Au, gives a large $H$-linear contribution to $\rho_{xy}$ at low temperatures.

It is very unlikely that the small isoelectronic $x=0.01$ and $0.03$ Ag (i.e. 0.2 and 0.6% substitution), $x=0.01$ Au, and $x=0.02$ Lu substitutions are significantly changing the overall electronic structure, and hence should not significantly change the contribution to the Berry phase from spin-orbit induced momentum-space magnetic monopoles in the band structure. Furthermore, the magnetic monopole induced Berry phase contribution to the anomalous Hall conductivity in a given material is intrinsic and expected to be independent of changes in the scattering rate due to disorder[41]. Hence, the suppression of the peak in $\rho_{xy}(H)$ with disorder argues against a large intrinsic (K-L) Berry phase contribution, and the failure of the $\rho_{xx}(T)$ data for the Au $x$=0.03 sample to show a sign of entering into M2 shows that the chirality-induced Berry phase due to the 4-q order in UCu$_5$ is necessary for the peak in the observed anomalous Hall response.

We do not know the mechanism by which these substitutions are influencing the 4-q magnetic structure, but it is apparent from these results that the 4-q structure is a prerequisite for the anomalous Hall effect in UCu$_5$ and that the 4-q state is extremely sensitive to disorder. In addition to the fact that we can model the peak in $\rho_{xy}(H)$ using the chirality of the spin texture, we take these results from substitution measurements as strong support for our conclusion that the peak in $\rho_{xy}(H)$ at $H$~2-2.5T for UCu$_5$ is due to a geometrical Hall effect arising from the chirality-induced Berry phase acquired as an electron traverses the non-coplanar 4-q magnetic order.

**Supplementary References**